\documentclass[%
 reprint,
superscriptaddress,
 amsmath,amssymb,
prb,
]{revtex4-1} 

\usepackage{graphicx}
\usepackage{dcolumn}
\usepackage{bm}
\usepackage[usenames,dvipsnames]{color}

\begin{document}

\preprint{APS/123-QED}

\title{Active control of thermal emission by graphene-nanowire coupled plasmonic metasurfaces}

\author{Jiayu Li}
\thanks{These two authors contribute equally.}
 \affiliation{Department of Mechanical Engineering, Carnegie Mellon University, Pittsburgh, Pennsylvania 15213, USA}

\author{Zhuo Li}
\thanks{These two authors contribute equally.}
 \affiliation{Department of Mechanical Engineering, Carnegie Mellon University, Pittsburgh, Pennsylvania 15213, USA}

 \author{Xiu Liu}
 \affiliation{Department of Mechanical Engineering, Carnegie Mellon University, Pittsburgh, Pennsylvania 15213, USA}
 
 \author{Stanislav Maslovski}
 \affiliation{Instituto de Telecomunica\c{c}\~{o}es and Department of Electronics, Telecommunications and Informatics, University of Aveiro, Campus Universit\'{a}rio de Santiago, 3810-193 Aveiro, Portugal}

 \author{Sheng Shen}
 \email{sshen1@cmu.edu}
 \affiliation{Department of Mechanical Engineering, Carnegie Mellon University, Pittsburgh, Pennsylvania 15213, USA}
 




\date{\today}

\begin{abstract}
 Metasurfaces, together with graphene plasmonics, have become prominent for the emissivity control in thermal engineering, both passively through changing the geometric parameters and packing density of the metasurfaces, and actively through graphene gating or doping. We demonstrate a graphene-nanowire coupled plasmonic metasurface utilizing the hybrid localized surface plasmon modes of the nanowire array and graphene. The nanowire array makes the hybrid surface plasmon mode localized, allowing a free-space excitation. The single layer graphene, via the gating between the underneath mirror and a top electrode, can actively tune the spectral emissivity by almost 90\%. In addition, the graphene surface plasmon modes remove the strict polarization dependence of nanowire array emission, resulting in a five-fold enhancement of the p-polarized emissivity, especially for large emission angles.
\end{abstract}

\maketitle

\section{\label{sec:intro}Introduction}
Thermal radiation, which physically originates from the spontaneous emission of thermally induced random currents in materials, is fundamental in many modern applications, including biological and chemical sensing\cite{lochbaum_-chip_2017}, thermal imaging and camouflage\cite{li_structured_2018,salihoglu_graphene-based_2018}, energy conversion and harvesting\cite{bierman_enhanced_2016,fan_near-perfect_2020}, and radiative cooling\cite{raman_passive_2014}. Compared to the isotropic and incoherent thermal emission from bulk surfaces, nanostructure-based metasurfaces have been successfully applied to precisely control the emissivity both spectrally \cite{liu_resonant_2017, rev1-1, rev1-2, rev1-3} and spatially\cite{yu_directional_2019, rev1-4, rev1-5}. Based on the Purcell effect, the sub-wavelength nanostructure, which supports localized surface plasmon polaritons, can serve as an optical resonator to drastically modulate the response of the nanostructure at designed resonant frequencies. The coupling between these `meta-atoms', which is well described by a tight bonding model\cite{li_scale_2020} or coupled mode theory\cite{zhu_temporal_2013}, provides rich degrees of freedom for the design of a metasurface.

Recently, graphene plasmonics has attracted extensive attention to further boost the plasmonic effects in thermal radiation engineering\cite{koppens_graphene_2011,yan_damping_2013,brar_electronic_2015,guo_infrared_2017}. Graphene, a two-dimensional single layer of carbon atoms, can support surface plasmon polaritons with a stronger optical confinement as compared to conventional noble metals. It has been reported that graphene nanostructures enable to manipulate the light in a dimension on the order of $100$ times smaller than the free-space wavelength\cite{koppens_graphene_2011}. This excellent beyond-diffraction-limit performance plays a critical role in bridging the nanoscale electronic devices and microscale photonic devices. In addition, the graphene plasmons excited in the mid-infrared range typically exhibit low loss, resulting in a stronger spectral coherence for thermal radiation\cite{jablan_plasmonics_2009}. More importantly, the optical responses of graphene can be actively tuned by changing its charge carrier density via gating or doping\cite{chen_controlling_2011,efetov_controlling_2010}. Together with the improving maturity of graphene transfer and patterning, graphene plasmonic metasurfaces become a promising complement to traditional metal plasmonics and pave the way for the ultra-fast control of thermal radiation.

\begin{figure}
\includegraphics[width = 3 in]{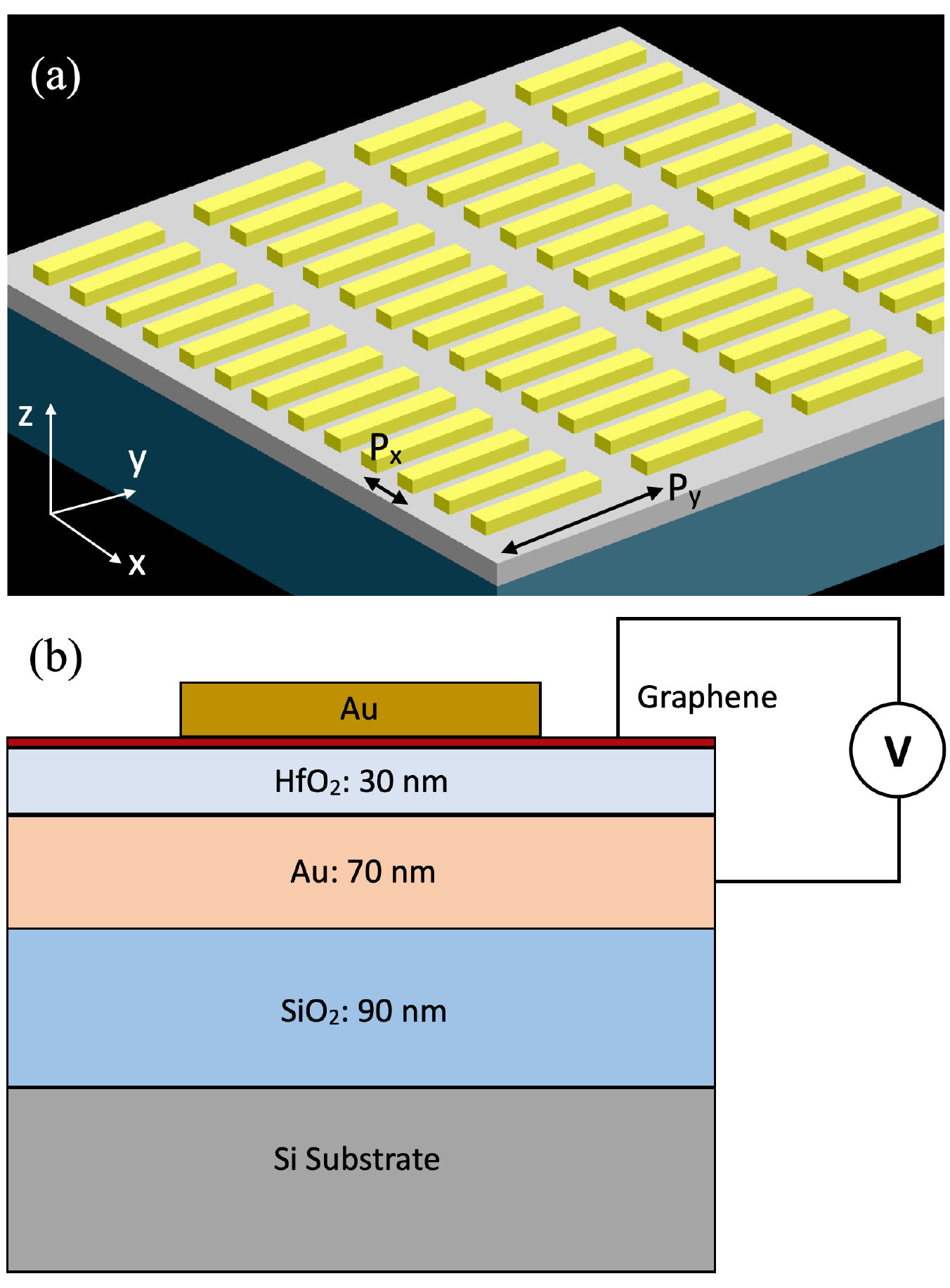}
\caption{\label{fig1}(a) Schematic of graphene-nanowire coupled metasurface. $P_x$ and $P_y$ represent the periodicity of the nanowire array in the $x$-and $y$-directions, respectively.  (b) Cross-section of the graphene-nanowire coupled metasurface. Electrical gating of graphene is achieved via the underneath Au mirror and the top electrode.}
\end{figure}

In this paper, we propose a graphene-nanowire coupled metasurface working at mid-infrared wavelengths, as shown in Figs.~\ref{fig1}~(a) and (b). The graphene layer is configured below the nanowire array to minimize undesired wrinkles, making our structure more feasible to fabricate than the cases with graphene on the top\cite{rev2-1,rev2-2, rev2-3, rev2-4, rev2-5}. The nanowire array makes the hybrid surface plasmon mode localized, allowing for a free-space excitation. The single layer graphene, via the gating between the underneath mirror and a top electrode, can actively tune the thermal emission. Compared to previous studies regarding gated graphene metasurfaces\cite{rev2-6, rev2-7, rev2-8}, we observe the resonance mode split (Sec. \ref{sec: absp mod}) and the P-polarization excitation (Sec. \ref{sec: p-polar enhancement}), which provide extra designing spaces for metasurfaces.

\section{\label{sec: system}Mode hybridization of Graphene - Nanowire Coupled Metasurface}
To demonstrate the physics associated with the graphene-nanowire hybrid surface plasmon mode, we perform finite-difference time-domain (FDTD) simulations on both a single nanowire and a nanowire array. The FDTD simulations are achieved via the Wiener chaos expansion (WCE) method\cite{li_tunable_2017,li_wiener_2021}, where a series of dipoles are assigned to the nanowire to systematically excite the supported modes. Detailed setting of boundary conditions of the simulations can be found in Appendix \ref{apdx: simulation setup}.

The gold nanowire (or nanowire array) is placed directly on top of the single graphene layer [Fig.\ref{fig1} (a)] (permittivity acquired from \cite{graphene_mat}). A $30~{\rm nm}$ thick layer of ${\rm HfO_2}$ (permittivity acquired from \cite{HfO2_mat}) is under the graphene acting as the dielectric spacer for electrical gating. Another $70~{\rm nm}$ thick gold (permittivity acquired from \cite{Au_mat}) layer under the ${\rm HfO_2}$ layer serves as an optical mirror as well as the electrical gating electrode. Then the layered structure is supported by $90~{\rm nm}$ ${\rm SiO_2}$ (permittivity acquired from \cite{Palik}) on top of a silicon substrate (permittivity acquired from \cite{Si_mat_1,Si_mat_2,Si_mat_3}),  as shown in Fig.~\ref{fig1}~(b). The sizes of the single nanowire are optimized to be $150~{\rm nm}$ ($x$-direction) by $2.5~{\rm \mu m}$ ($y$-direction) by $180~{\rm nm}$ ($z$-direction); as well as the periodicities of the nanowire array $P_x = 0.5~{\rm \mu m}$ and $P_y = 3~{\rm \mu m}$. The optimization criterion and processes are detailed in Appendix \ref{apdx: opt_of_parameters}.

Figure \ref{fig2} (a) shows the first direct emission resonant mode of a single nanowire on top of the intrinsic graphene layer. It manifests a typical dipole-like resonance characteristic with charges accumulated at the two ends of the nanowire. Without gating, only the localized plasmon mode from the gold nanowire can be observed, since the intrinsic graphene performs like a dielectric layer. 
\begin{figure}
\includegraphics[width = 3.1 in]{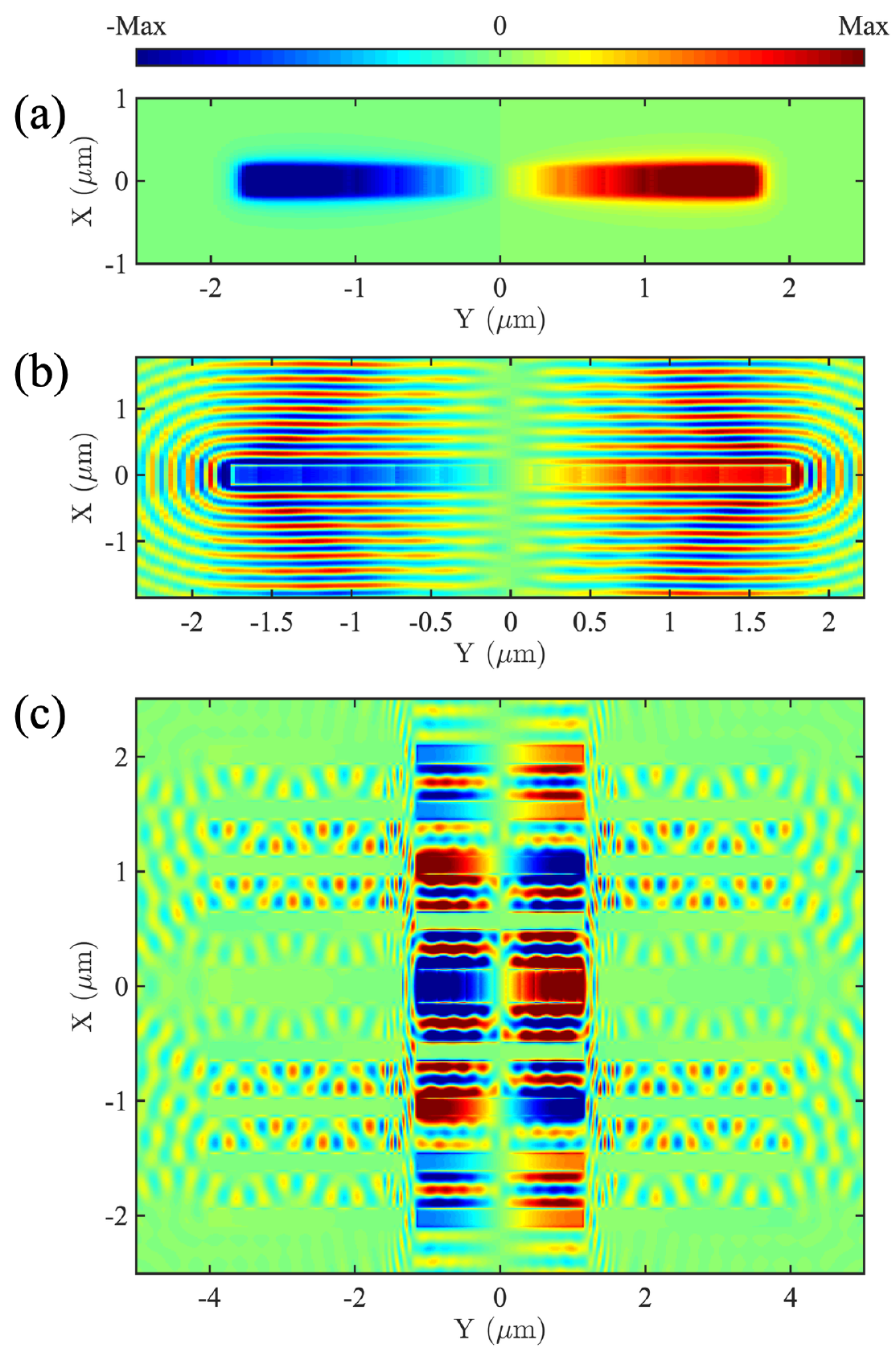}
\caption{\label{fig2} $E_z$ field profiles at $\lambda = 10.8 {\rm \mu m}$ that illustrate (a) the fundamental resonant mode of the single gold nanowire on the intrinsic graphene layer, (b) the hybridized surface plasmon mode excited within the single nanowire on the gated graphene layer with the Fermi level at $0.4~{\rm eV}$, and (c) the hybridized surface plasmon mode excited within the nanowire array on the gated graphene layer with Fermi level at $0.4~{\rm eV}$. Field profiles captured at the graphene-nanowire (array) interface. The color scale represents the field magnitude.}
\end{figure}

By gating the graphene layer, it behaves like a metallic material due to the extra free electrons, and thus surface plasmon modes can be excited at the metal-graphene interface. The single nanowire then functions as the near-field scattering tip to efficiently excite the surface plasmon modes along the graphene layer. As we excite the graphene-nanowire system, not only the nanowire resonant modes but also the graphene surface plasmon can be excited, as shown in Fig.~\ref{fig2}~(b). With the single nanowire deposited on the graphene layer functioning as the scattering tip, the surface plasmon modes supported by the graphene layer will continuously extend to the entire graphene layer until their energy completely decays.

However, if an array of nanowires is introduced and forms a metasurface above the graphene layer, the continuous surface plasmon modes supported by graphene will then be transformed into the localized ones. As one can see from Fig.~\ref{fig2}~(c), the graphene surface plasmon modes are excited and trapped by the contact points between the nanowires and graphene layer. Such hybrid surface plasmon modes are the key to achieve active modulation of the optical properties (Sec. \ref{sec: absp mod}) of the metasurface via gating the graphene layer, which is a pure electrical process. Therefore, the modulation speed is mainly limited by the electrical RC time constant of which the typical number is on the order of 10 GHz\cite{xiu-1,xiu-2,xiu-3}, and is much faster than the speed of thermal modulation, typically less than 20MHz\cite{xiu-4}.

\section{\label{sec: absp mod} Emissivity modulation of the graphene-nanowire metasurface}
\begin{figure*}
\includegraphics[width = 5.5 in]{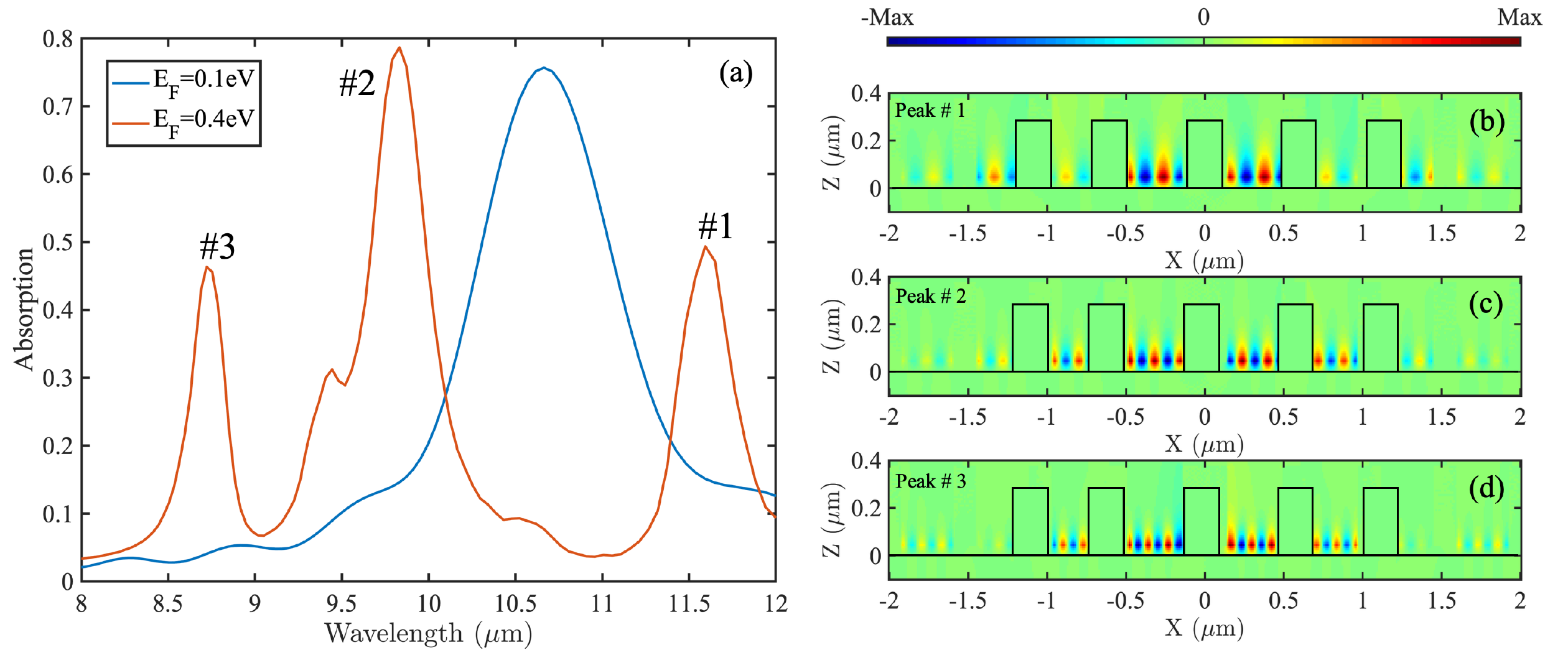}
\caption{\label{fig3}(a) The modulation of the absorption (emission) spectrum for the graphene-nanowire metasurface. Nearly 90\% of the change in the absorption spectrum around 10.5~$\mu$m in wavelength is observed when the Fermi level of graphene is changed from $0.1~{\rm eV}$ to $0.4~{\rm eV}$ due to the mode splitting. (b)--(d) $E_x$ field profiles corresponding to peaks \#1 to \#3, respectively. The field profiles are captured in the $x-z$ plane $1 {\rm \mu m}$ away from the center of nanowires. The color scale represents the field magnitude.}
\end{figure*}

Here we demonstrate the active modulation of the emissivity of the graphene-nanowire coupled metasurface. The parameters of the nanowire array are the same as used in Sec. \ref{sec: system}, and the Fermi level of graphene is gated to be 0.1 or 0.4 ${\rm eV}$. The emissivity of the system is calculated by simulating its absorptivity to the plane wave polarized along the principal axis of the nanowire, based on Kirchhoff's law. The corresponding metasurface emissivity can then be found to dramatically change from 0.05 to 0.8 at the wavelength about $10.5~{\rm \mu m}$ when the Fermi level of the graphene is tuned from $0.4~{\rm eV}$ to $0.1~{\rm eV}$, as shown in Fig.~\ref{fig3}~(a).

Such a control of the emissivity at a certain wavelength is achieved based on the shifting and splitting of the nanowire resonant modes. As shown in Fig.~\ref{fig2}~(c), the hybrid plasmonic mode is formed by two main contributions: the intrinsic resonant modes of the gold nanowires themselves; and the localized surface plasmon modes existing inside the cavity created by the adjacent nanowires and the graphene layer. When the graphene layer is not gated, no surface plasmon mode is supported, and the original resonant peak of the gold nanowire is preserved at $10.5~{\rm \mu m}$. When the graphene layer is electrically gated, it becomes metallic and starts to support the surface plasmon modes. The first resonant mode of the gold nanowire hybridizes with the graphene surface plasmon modes of different orders and splits into multiple peaks around the original peak wavelength. 

As shown in Fig.~\ref{fig3}~(a), when the Fermi level is $0.1~{\rm eV}$, the carrier density in graphene is relatively low and hence surface plasmon modes are not supported in graphene. The original nanowire resonant mode peak is still preserved at $10.5~{\rm \mu m}$. As we increase the Fermi level up to $0.4~{\rm eV}$, the major resonant peak at $10.5~{\rm \mu m}$ splits into three peaks at $8.7~{\rm \mu m}$, $9.7~{\rm \mu m}$, and $11.7~{\rm \mu m}$ due to hybridization of the plasmon modes in the gold nanowire and in graphene. The corresponding electric field profiles of these three peaks are plotted in Figs.~\ref{fig3}~(b)--(d). Different spatial periodicities of the electric field are observed in the three cases and label the order of the modes, among which peak \#1 is the surface plasmon mode of the lowest order, and peak \#3 is the highest order mode in the wavelength range of interest. The number of the hybridized peaks around the original nanowire resonant modes and the spectral distance between them can be controlled by the gap distance between the adjacent nanowires.
Ideally, carefully designing the pattern periodicity and the gold nanowire parameters can shift the desired wavelength to a target value in the mid-infrared spectrum.

\section{\label{sec: p-polar enhancement}p-Polarization enhancement of the Graphene-Nanowire metasurface}

\begin{figure*}
\includegraphics[width = 6.1 in]{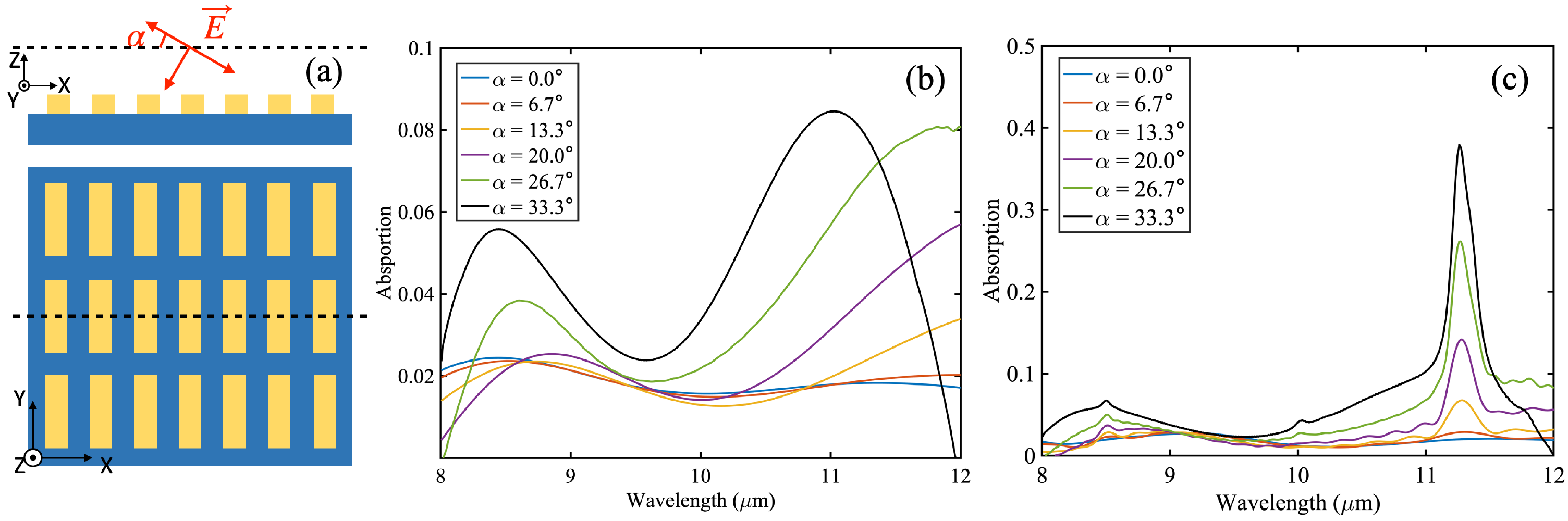}
\caption{\label{fig4}  (a) Schematic of the oblique incidence of the P-polarized light onto the metasurface at angle $\alpha$. The periodicity of the nanowire array is $0.3~{\rm \mu m}$ in the $x$-direction and $3~{\rm \mu m}$ in the $y$-direction. The angular dependent spectral emissivity of the metasurface when the graphene Fermi level is tuned to be (b) $0~{\rm eV}$, and (c) $0.6~{\rm eV}$.}
\end{figure*}

Due to the highly polarized dipole-like radiation mode of the single nanowire, the nanowire metasurface usually exhibits a low emissivity when excited by the p-polarized electromagnetic wave [i.e., the E-field polarization is perpendicular to the principal axis of the nanowires, as shown in Fig.~\ref{fig4}~(a)]. Here we demonstrate that such a P-polarized emissivity can be significantly enhanced for the graphene-nanowire array system when graphene is gated, especially for large incident angles, as indicated by $\alpha$ in Fig.~\ref{fig4}~(a).

Figure \ref{fig4}~(b) plots the emissivity of the metasurface when the Fermi level of graphene is set to $0~{\rm eV}$. In this case, the incoming plane wave source cannot efficiently excite the metasurface due to the polarization mismatch, which results in the relatively low emissivity. Despite the overall low value, a peak centered around $8.5~{\rm \mu m}$ can be noticed for the cases where $\alpha = 26.7^{\circ}$ and $\alpha = 33.3^{\circ}$, and another centered around $11~{\rm \mu m}$ can be noticed for $\alpha = 33.3 ^{\circ}$, which correspond to the surface plasmon modes  of the graphene excited by the scattered light from nanowires and can be utilized to enhance the p-polarized emissivity of the system. 

To achieve this, we electronically gate the graphene layer so that its Fermi level reaches $0.6~{\rm eV}$. The simulated spectral emissivities are plotted in Fig.~\ref{fig4}~(c). A narrow peak centered at the wavelength of $11.3~{\rm \mu m}$ is observed for the incident angle $\alpha$ of the light between $13.3^{\circ}$ to $33.3^{\circ}$. Specifically, compared to the ungated cases shown in Fig.~\ref{fig4}~(b), a five-fold increase in the emissivity at the wavelength of $11.3~{\rm \mu m}$ is observed for the cases where the incident angle $\alpha$ is between $20.0^{\circ}$ to $33.3^{\circ}$.

Such enhancement of the p-polarized emissivity is attributed to the cross polarization excitation of the nanowire modes with the assistance of the graphene surface plasmon modes. Even though the p-polarized wave cannot effectively excite the resonant modes of the nanowire array, the scattered light from the nanowires will be able to excite the surface plasmon modes inside the graphene layer, especially for the large incident angle $\alpha$. The surface plasmon modes of the graphene may then serve as a secondary source to excite the dipole-mode of nanowires, which produces the sharp peak near the intrinsic resonance frequency of nanowires.

\begin{figure*}
\includegraphics[width = 6.1 in]{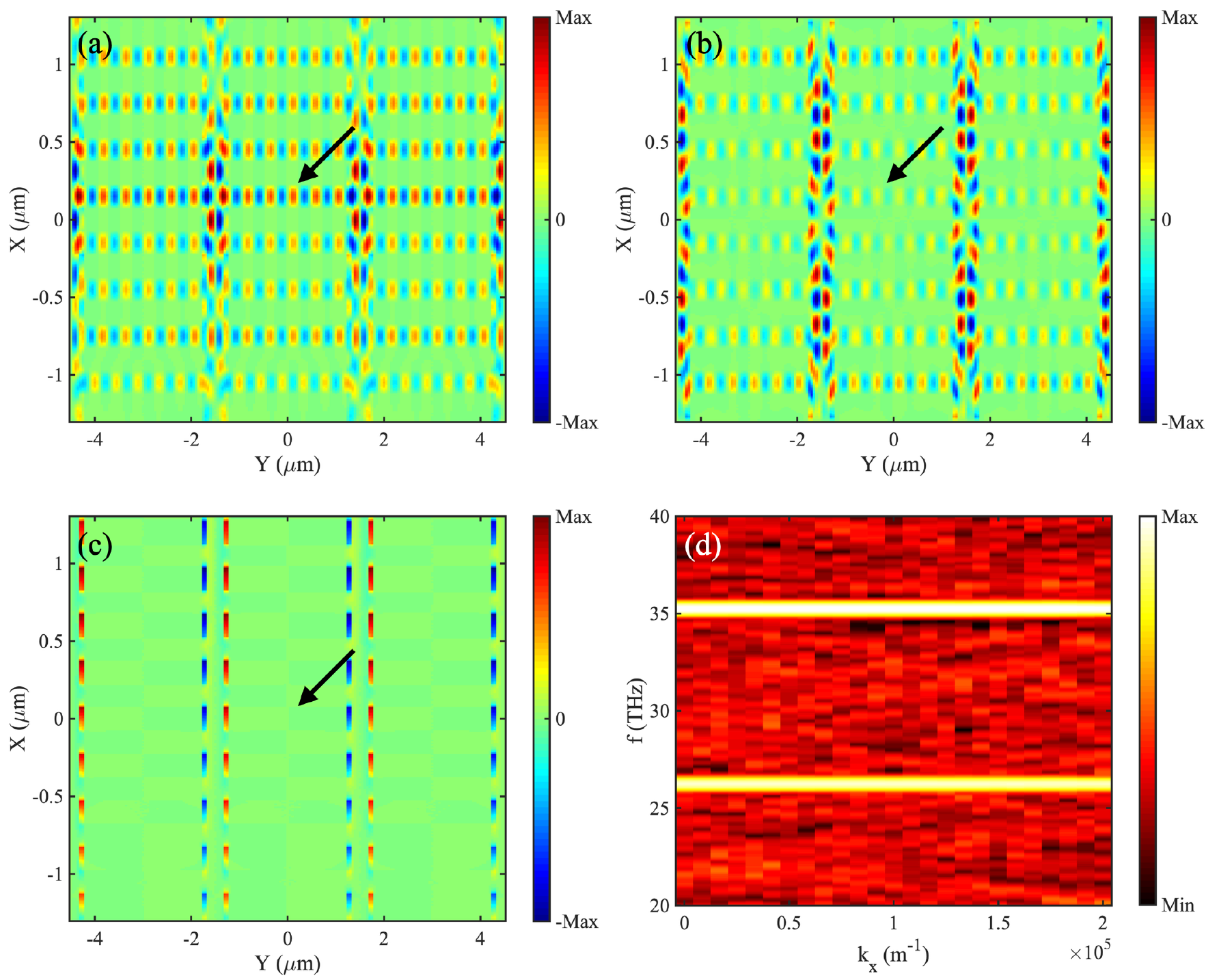}
\caption{\label{fig5}Simulated $E_x$ profile at $\lambda=\ 11.3{\rm \mu m}$ at the graphene-nanowire array interface when the metasurface is illuminated by the x-polarized plane wave under the conditions: (a) $E_F = 0.6~{\rm  eV}$, $\alpha = 33.3^{\circ}$, (b) $E_F = 0.6~{\rm  eV}$, $\alpha = 0^{\circ}$, and (c) $E_F = 0~{\rm  eV}$, $\alpha = 33.3^{\circ}$. Black arrows highlight the graphene surface plasmon modes, which serve as the secondary source to excite the nanowire array. (d) Simulated dispersion relation (colormap in the log scale) when $E_F = 0.6~{\rm eV}$.}
\end{figure*}

The Fermi level of graphene and the incident direction of the source play a key role in a successful cross polarization excitation. In Fig. \ref{fig5}, we compare the field profiles at the nanowire-graphene interface at different conditions. Compared to the successful excitation [Fig.\ref{fig5} (a)], the surface plasmon modes of graphene in the $y$-direction are not pronounced when the incoming light is normally incident [Fig.\ref{fig5} (b)] due to the varnished $k_{\parallel}$ of the incoming light, and when the Fermi level of graphene is 0 eV [Fig.\ref{fig5} (c)] due to the lack of free carriers to support the plasmon mode. Hence the cross polarization excitation is not achieved, which results in the low absorption.

Figure \ref{fig5} (d) shows the dispersion relations of the nanowire array, coupled with graphene with the Fermi level of $0.6~{\rm eV}$, with respect to $k_x$. It confirms the existence of the mode excited: the bottom bright line represents the peak (mode) around  $11.3~ {\rm \mu m}$, whereas the top bright line agrees with the less-pronounced mode around $8.5~ {\rm \mu m}$, which may correspond to the high-order mode of the nanowire array resonance.

Such a cross polarization excitation enhances the P-polarization absorption of nanowire arrays, and relieves their strict dependence on the polarization of the incident light. Furthermore, the Fermi level of graphene, together with the incident angle $\alpha$, provides extra degrees of freedom for the thermal radiation manipulation of the graphene-nanowire array system.

\section{\label{sec: conclusion}Conclusion}
We develop a graphene-nanowire coupled metasurface utilizing the hybrid localized surface plasmon modes of the nanowire array and graphene. Thermal excitation can excite both the resonant modes of the nanowire array and the graphene surface plasmon modes trapped at the contact points between the nanowire and the graphene layer. These coupled modes can drastically tune the emissivity of the metasurface passively through changing the geometric parameters and packing density of the nanowire array, and more importantly, actively through graphene gating. The peak positions and corresponding magnitudes of spectral emissivity can be modulated into target values through controlling the graphene Fermi level via electrical gating. In addition, with the assistance of graphene surface plasmon modes, the P-polarized emissivity of the metasurface can be enhanced, especially for large emission angles. The graphene-coupled nanowire metasurface is a promising platform for dynamic control of mid-infrared thermal radiation.

The data that support the findings of this study are available from the corresponding author upon request\footnote{The data that support the findings of this study are available from the corresponding author upon request}.

\begin{acknowledgments}
This work was primarily supported by the Defense Threat Reduction Agency (Grant No. HDTRA1-19-1-0028). This work was also funded partially by the National Science Foundation (Grant No. CBET-1931964). J.~L. and Z.~L. contribute equally. S.~S. and S.~M.\ acknowledge funding from Funda\c{c}\~{a}o para a Ci\^{e}ncia e a Tecnologia (FCT), Portugal, under the Carnegie Mellon Portugal Program (project ref. CMU/TIC/0080/2019). S.~M. also acknowledges support received from FCT/MCTES through national funds and when applicable co-funded from the EU funds under the project UIDB/50008/2020-UIDP/50008/2020.

The authors declare no competing financial interest.
\end{acknowledgments}

\appendix
\section{Simulation Setup}
\label{apdx: simulation setup}

\begin{figure}
\includegraphics[width = 2.5 in]{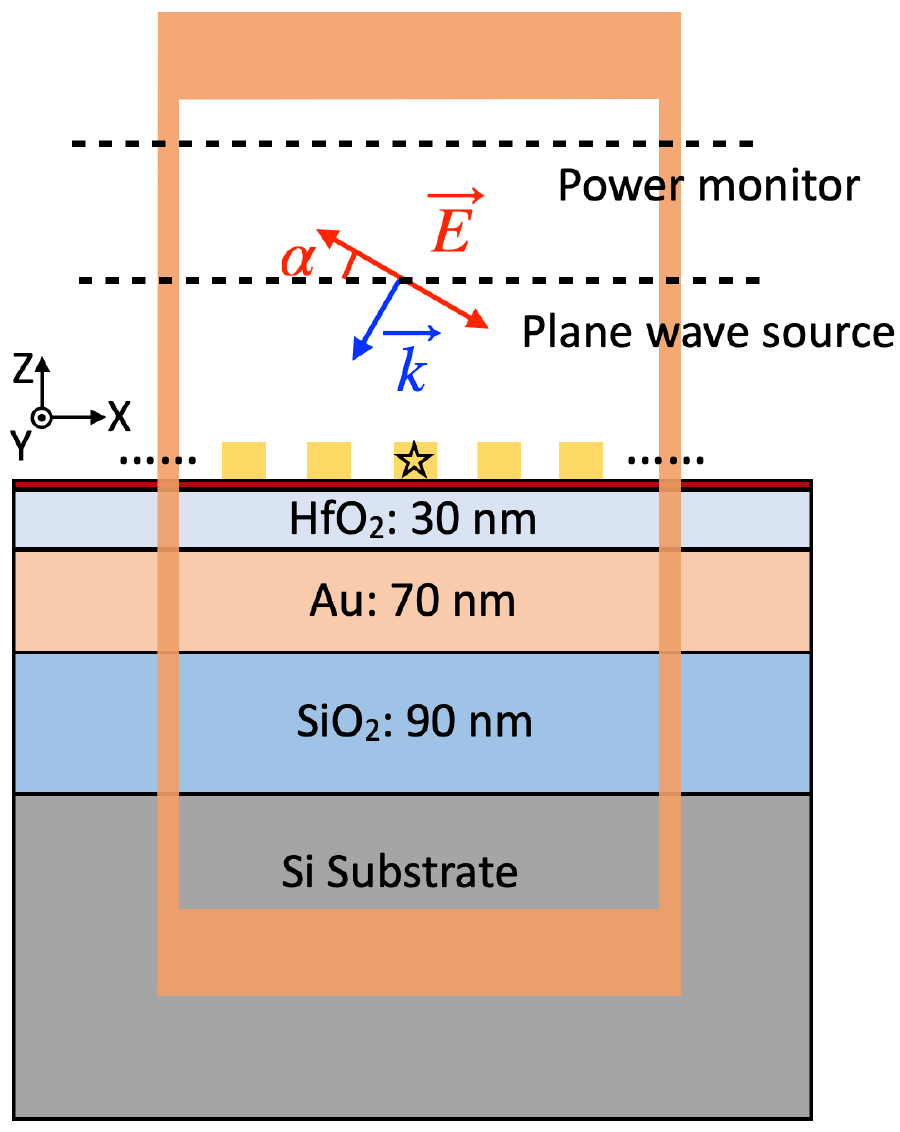}
\caption{\label{figA1} Simulation setup in Ansys Lumerical FDTD Solutions. Orange Boxes mark the positions of simulation boundaries, where the boundary conditions are applied.}
\end{figure}

The simulation setup is shown in Fig. \ref{figA1}. The orange box marks the boundaries of the simulation domains. Through the entire study, the boundary conditions (BCs) at the boundaries in the $z$-direction are set to be perfectly matched layers (PMLs).  The BCs in the lateral directions are detailed as follows.

In Fig. \ref{fig2}, only the nanowire located at the center of the simulation domain (marked by the black star) is excited by dipole sources via the WCE method, and the simulation boundaries in the $x$ and $y$ directions are set to be PMLs.  Therefore, the field profile shown in Fig. \ref{fig2}(c) corresponds to a finite-sized array.

In Figs. \ref{fig3} and \ref{fig4}, infinite arrays illuminated by plane wave sources are simulated. This is achieved by setting the simulation boundaries in the $x$ and $y$ directions to be periodic boundaries, with the exceptions for oblique incidence cases discussed in Fig. \ref{fig4}, in which the simulation boundaries in the x direction are set to Bloch boundaries.

\section{Optimization of nanowire array parameters}
\label{apdx: opt_of_parameters}

The sizes and periodicities of the nanowire arrays studied are optimized via a process similar to that used in\cite{li_scale_2020, yu_directional_2019}, and achieved by the parameter sweep function in Ansys Lumerical FDTD Solutions.

\begin{figure}
\includegraphics[width = 3 in]{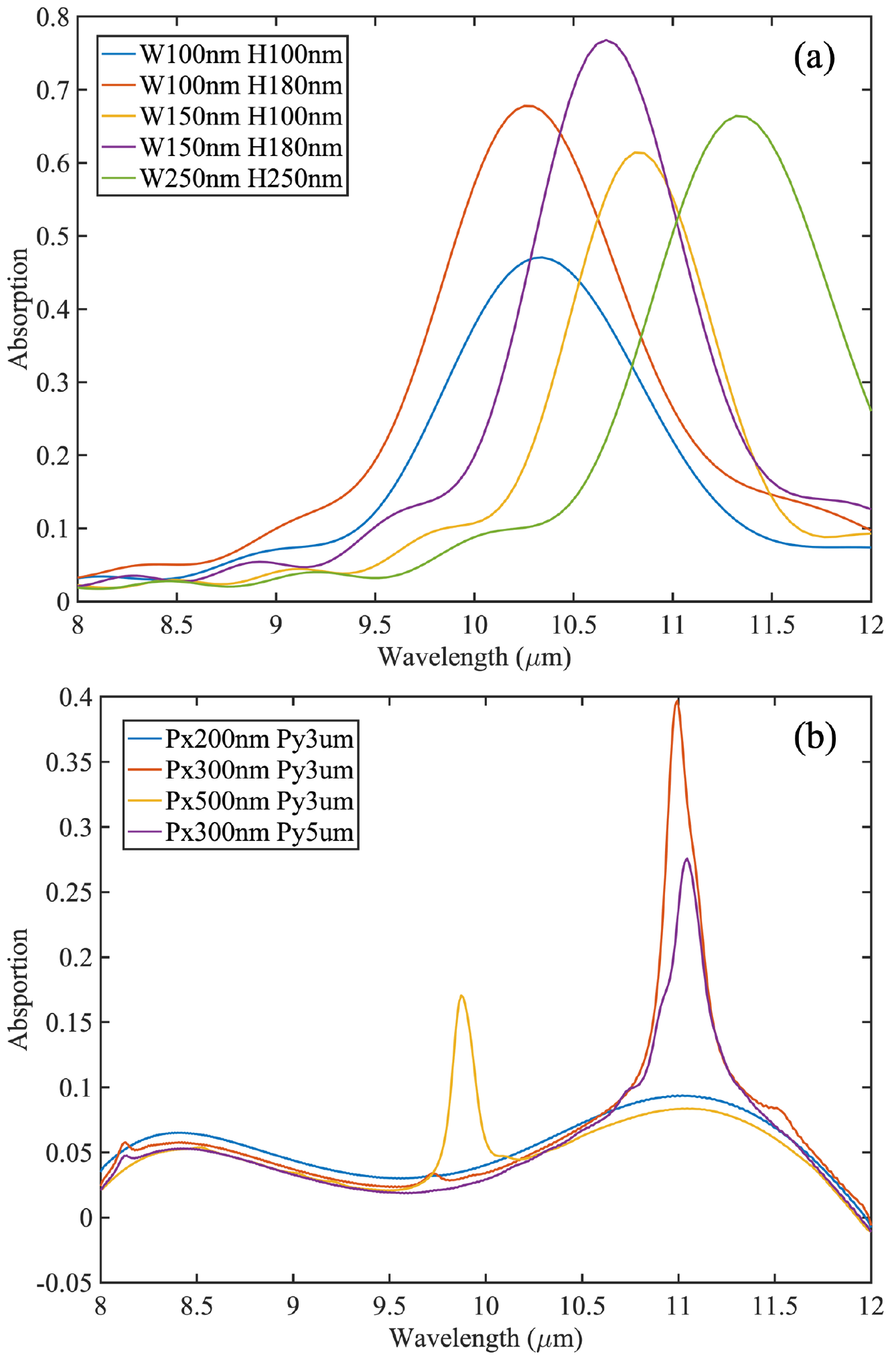}
\caption{\label{figA2}  Optimization process for (a) width and thickness of nanowires, and (b) periodicities of nanowire array in the $x-$ and $y-$ directions.}
\end{figure}

The optimization criterion for the width and thickness of nanowires is to maximize the absorption peak when the array is on top of the graphene sheet with $E_f=0.1\ {\rm eV}$ and excited by the light polarized along the nanowire. Optimized combination of 150 nm in width and 180 nm in thickness is chosen, as shown in Fig. \ref{figA2} (a) (here the periodicities in the $x$ and $y$ directions are fixed to be 500 nm and $3~{\rm \mu m}$, respectively).

The optimization criterion for the periodicities is to maximize the P-polarization enhancement, discussed in Sec. \ref{sec: p-polar enhancement}. As shown in Fig. \ref{figA2} (b), the optimized periodicity combination ($P_x = 300~{\rm nm}$ and $Py =  3\ {\rm \mu m}$) is used for this study.

\clearpage


\begin{thebibliography}{44}%
\makeatletter
\providecommand \@ifxundefined [1]{%
 \@ifx{#1\undefined}
}%
\providecommand \@ifnum [1]{%
 \ifnum #1\expandafter \@firstoftwo
 \else \expandafter \@secondoftwo
 \fi
}%
\providecommand \@ifx [1]{%
 \ifx #1\expandafter \@firstoftwo
 \else \expandafter \@secondoftwo
 \fi
}%
\providecommand \natexlab [1]{#1}%
\providecommand \enquote  [1]{``#1''}%
\providecommand \bibnamefont  [1]{#1}%
\providecommand \bibfnamefont [1]{#1}%
\providecommand \citenamefont [1]{#1}%
\providecommand \href@noop [0]{\@secondoftwo}%
\providecommand \href [0]{\begingroup \@sanitize@url \@href}%
\providecommand \@href[1]{\@@startlink{#1}\@@href}%
\providecommand \@@href[1]{\endgroup#1\@@endlink}%
\providecommand \@sanitize@url [0]{\catcode `\\12\catcode `\$12\catcode
  `\&12\catcode `\#12\catcode `\^12\catcode `\_12\catcode `\%12\relax}%
\providecommand \@@startlink[1]{}%
\providecommand \@@endlink[0]{}%
\providecommand \url  [0]{\begingroup\@sanitize@url \@url }%
\providecommand \@url [1]{\endgroup\@href {#1}{\urlprefix }}%
\providecommand \urlprefix  [0]{URL }%
\providecommand \Eprint [0]{\href }%
\providecommand \doibase [0]{http://dx.doi.org/}%
\providecommand \selectlanguage [0]{\@gobble}%
\providecommand \bibinfo  [0]{\@secondoftwo}%
\providecommand \bibfield  [0]{\@secondoftwo}%
\providecommand \translation [1]{[#1]}%
\providecommand \BibitemOpen [0]{}%
\providecommand \bibitemStop [0]{}%
\providecommand \bibitemNoStop [0]{.\EOS\space}%
\providecommand \EOS [0]{\spacefactor3000\relax}%
\providecommand \BibitemShut  [1]{\csname bibitem#1\endcsname}%
\let\auto@bib@innerbib\@empty
\bibitem [{\citenamefont {Lochbaum}\ \emph {et~al.}(2017)\citenamefont
  {Lochbaum}, \citenamefont {Fedoryshyn}, \citenamefont {Dorodnyy},
  \citenamefont {Koch}, \citenamefont {Hafner},\ and\ \citenamefont
  {Leuthold}}]{lochbaum_-chip_2017}%
  \BibitemOpen
  \bibfield  {author} {\bibinfo {author} {\bibfnamefont {A.}~\bibnamefont
  {Lochbaum}}, \bibinfo {author} {\bibfnamefont {Y.}~\bibnamefont
  {Fedoryshyn}}, \bibinfo {author} {\bibfnamefont {A.}~\bibnamefont
  {Dorodnyy}}, \bibinfo {author} {\bibfnamefont {U.}~\bibnamefont {Koch}},
  \bibinfo {author} {\bibfnamefont {C.}~\bibnamefont {Hafner}}, \ and\ \bibinfo
  {author} {\bibfnamefont {J.}~\bibnamefont {Leuthold}},\ }\href {\doibase
  10.1021/acsphotonics.6b01025} {\bibfield  {journal} {\bibinfo  {journal} {ACS
  Photonics}\ }\textbf {\bibinfo {volume} {4}},\ \bibinfo {pages} {1371}
  (\bibinfo {year} {2017})}\BibitemShut {NoStop}%
\bibitem [{\citenamefont {Li}\ \emph {et~al.}(2018)\citenamefont {Li},
  \citenamefont {Bai}, \citenamefont {Yang}, \citenamefont {Luo},\ and\
  \citenamefont {Qiu}}]{li_structured_2018}%
  \BibitemOpen
  \bibfield  {author} {\bibinfo {author} {\bibfnamefont {Y.}~\bibnamefont
  {Li}}, \bibinfo {author} {\bibfnamefont {X.}~\bibnamefont {Bai}}, \bibinfo
  {author} {\bibfnamefont {T.}~\bibnamefont {Yang}}, \bibinfo {author}
  {\bibfnamefont {H.}~\bibnamefont {Luo}}, \ and\ \bibinfo {author}
  {\bibfnamefont {C.-W.}\ \bibnamefont {Qiu}},\ }\href {\doibase
  10.1038/s41467-017-02678-8} {\bibfield  {journal} {\bibinfo  {journal}
  {Nature Communications}\ }\textbf {\bibinfo {volume} {9}},\ \bibinfo {pages}
  {273} (\bibinfo {year} {2018})}\BibitemShut {NoStop}%
\bibitem [{\citenamefont {Salihoglu}\ \emph {et~al.}(2018)\citenamefont
  {Salihoglu}, \citenamefont {Uzlu}, \citenamefont {Yakar}, \citenamefont
  {Aas}, \citenamefont {Balci}, \citenamefont {Kakenov}, \citenamefont {Balci},
  \citenamefont {Olcum}, \citenamefont {Süzer},\ and\ \citenamefont
  {Kocabas}}]{salihoglu_graphene-based_2018}%
  \BibitemOpen
  \bibfield  {author} {\bibinfo {author} {\bibfnamefont {O.}~\bibnamefont
  {Salihoglu}}, \bibinfo {author} {\bibfnamefont {H.~B.}\ \bibnamefont {Uzlu}},
  \bibinfo {author} {\bibfnamefont {O.}~\bibnamefont {Yakar}}, \bibinfo
  {author} {\bibfnamefont {S.}~\bibnamefont {Aas}}, \bibinfo {author}
  {\bibfnamefont {O.}~\bibnamefont {Balci}}, \bibinfo {author} {\bibfnamefont
  {N.}~\bibnamefont {Kakenov}}, \bibinfo {author} {\bibfnamefont
  {S.}~\bibnamefont {Balci}}, \bibinfo {author} {\bibfnamefont
  {S.}~\bibnamefont {Olcum}}, \bibinfo {author} {\bibfnamefont
  {S.}~\bibnamefont {Süzer}}, \ and\ \bibinfo {author} {\bibfnamefont
  {C.}~\bibnamefont {Kocabas}},\ }\href {\doibase 10.1021/acs.nanolett.8b01746}
  {\bibfield  {journal} {\bibinfo  {journal} {Nano Letters}\ }\textbf {\bibinfo
  {volume} {18}},\ \bibinfo {pages} {4541} (\bibinfo {year}
  {2018})}\BibitemShut {NoStop}%
\bibitem [{\citenamefont {Bierman}\ \emph {et~al.}(2016)\citenamefont
  {Bierman}, \citenamefont {Lenert}, \citenamefont {Chan}, \citenamefont
  {Bhatia}, \citenamefont {Celanović}, \citenamefont {Soljačić},\ and\
  \citenamefont {Wang}}]{bierman_enhanced_2016}%
  \BibitemOpen
  \bibfield  {author} {\bibinfo {author} {\bibfnamefont {D.~M.}\ \bibnamefont
  {Bierman}}, \bibinfo {author} {\bibfnamefont {A.}~\bibnamefont {Lenert}},
  \bibinfo {author} {\bibfnamefont {W.~R.}\ \bibnamefont {Chan}}, \bibinfo
  {author} {\bibfnamefont {B.}~\bibnamefont {Bhatia}}, \bibinfo {author}
  {\bibfnamefont {I.}~\bibnamefont {Celanović}}, \bibinfo {author}
  {\bibfnamefont {M.}~\bibnamefont {Soljačić}}, \ and\ \bibinfo {author}
  {\bibfnamefont {E.~N.}\ \bibnamefont {Wang}},\ }\href {\doibase
  10.1038/nenergy.2016.68} {\bibfield  {journal} {\bibinfo  {journal} {Nature
  Energy}\ }\textbf {\bibinfo {volume} {1}},\ \bibinfo {pages} {1} (\bibinfo
  {year} {2016})}\BibitemShut {NoStop}%
\bibitem [{\citenamefont {Fan}\ \emph {et~al.}(2020)\citenamefont {Fan},
  \citenamefont {Burger}, \citenamefont {McSherry}, \citenamefont {Lee},
  \citenamefont {Lenert},\ and\ \citenamefont
  {Forrest}}]{fan_near-perfect_2020}%
  \BibitemOpen
  \bibfield  {author} {\bibinfo {author} {\bibfnamefont {D.}~\bibnamefont
  {Fan}}, \bibinfo {author} {\bibfnamefont {T.}~\bibnamefont {Burger}},
  \bibinfo {author} {\bibfnamefont {S.}~\bibnamefont {McSherry}}, \bibinfo
  {author} {\bibfnamefont {B.}~\bibnamefont {Lee}}, \bibinfo {author}
  {\bibfnamefont {A.}~\bibnamefont {Lenert}}, \ and\ \bibinfo {author}
  {\bibfnamefont {S.~R.}\ \bibnamefont {Forrest}},\ }\href {\doibase
  10.1038/s41586-020-2717-7} {\bibfield  {journal} {\bibinfo  {journal}
  {Nature}\ }\textbf {\bibinfo {volume} {586}},\ \bibinfo {pages} {237}
  (\bibinfo {year} {2020})}\BibitemShut {NoStop}%
\bibitem [{\citenamefont {Raman}\ \emph {et~al.}(2014)\citenamefont {Raman},
  \citenamefont {Anoma}, \citenamefont {Zhu}, \citenamefont {Rephaeli},\ and\
  \citenamefont {Fan}}]{raman_passive_2014}%
  \BibitemOpen
  \bibfield  {author} {\bibinfo {author} {\bibfnamefont {A.~P.}\ \bibnamefont
  {Raman}}, \bibinfo {author} {\bibfnamefont {M.~A.}\ \bibnamefont {Anoma}},
  \bibinfo {author} {\bibfnamefont {L.}~\bibnamefont {Zhu}}, \bibinfo {author}
  {\bibfnamefont {E.}~\bibnamefont {Rephaeli}}, \ and\ \bibinfo {author}
  {\bibfnamefont {S.}~\bibnamefont {Fan}},\ }\href {\doibase
  10.1038/nature13883} {\bibfield  {journal} {\bibinfo  {journal} {Nature}\
  }\textbf {\bibinfo {volume} {515}},\ \bibinfo {pages} {540} (\bibinfo {year}
  {2014})}\BibitemShut {NoStop}%
\bibitem [{\citenamefont {Liu}\ \emph {et~al.}(2017)\citenamefont {Liu},
  \citenamefont {Li},\ and\ \citenamefont {Shen}}]{liu_resonant_2017}%
  \BibitemOpen
  \bibfield  {author} {\bibinfo {author} {\bibfnamefont {B.}~\bibnamefont
  {Liu}}, \bibinfo {author} {\bibfnamefont {J.}~\bibnamefont {Li}}, \ and\
  \bibinfo {author} {\bibfnamefont {S.}~\bibnamefont {Shen}},\ }\href {\doibase
  10.1021/acsphotonics.7b00336} {\bibfield  {journal} {\bibinfo  {journal} {ACS
  Photonics}\ }\textbf {\bibinfo {volume} {4}},\ \bibinfo {pages} {1552}
  (\bibinfo {year} {2017})}\BibitemShut {NoStop}%
\bibitem [{\citenamefont {Pralle}\ \emph {et~al.}(2002)\citenamefont {Pralle},
  \citenamefont {Moelders}, \citenamefont {McNeal}, \citenamefont {Puscasu},
  \citenamefont {Greenwald}, \citenamefont {Daly}, \citenamefont {Johnson},
  \citenamefont {George}, \citenamefont {Choi}, \citenamefont {El-Kady},\ and\
  \citenamefont {Biswas}}]{rev1-1}%
  \BibitemOpen
  \bibfield  {author} {\bibinfo {author} {\bibfnamefont {M.~U.}\ \bibnamefont
  {Pralle}}, \bibinfo {author} {\bibfnamefont {N.}~\bibnamefont {Moelders}},
  \bibinfo {author} {\bibfnamefont {M.~P.}\ \bibnamefont {McNeal}}, \bibinfo
  {author} {\bibfnamefont {I.}~\bibnamefont {Puscasu}}, \bibinfo {author}
  {\bibfnamefont {A.~C.}\ \bibnamefont {Greenwald}}, \bibinfo {author}
  {\bibfnamefont {J.~T.}\ \bibnamefont {Daly}}, \bibinfo {author}
  {\bibfnamefont {E.~A.}\ \bibnamefont {Johnson}}, \bibinfo {author}
  {\bibfnamefont {T.}~\bibnamefont {George}}, \bibinfo {author} {\bibfnamefont
  {D.~S.}\ \bibnamefont {Choi}}, \bibinfo {author} {\bibfnamefont
  {I.}~\bibnamefont {El-Kady}}, \ and\ \bibinfo {author} {\bibfnamefont
  {R.}~\bibnamefont {Biswas}},\ }\href {\doibase 10.1063/1.1526919} {\bibfield
  {journal} {\bibinfo  {journal} {Applied Physics Letters}\ }\textbf {\bibinfo
  {volume} {81}},\ \bibinfo {pages} {4685} (\bibinfo {year} {2002})},\ \Eprint
  {http://arxiv.org/abs/https://doi.org/10.1063/1.1526919}
  {https://doi.org/10.1063/1.1526919} \BibitemShut {NoStop}%
\bibitem [{\citenamefont {Iida}\ \emph {et~al.}(2004)\citenamefont {Iida},
  \citenamefont {Tani}, \citenamefont {Sakai}, \citenamefont {Watanabe},
  \citenamefont {Katayama}, \citenamefont {Kondo}, \citenamefont {Kitahara},
  \citenamefont {Kato},\ and\ \citenamefont {Takeda}}]{rev1-2}%
  \BibitemOpen
  \bibfield  {author} {\bibinfo {author} {\bibfnamefont {M.}~\bibnamefont
  {Iida}}, \bibinfo {author} {\bibfnamefont {M.}~\bibnamefont {Tani}}, \bibinfo
  {author} {\bibfnamefont {K.}~\bibnamefont {Sakai}}, \bibinfo {author}
  {\bibfnamefont {M.}~\bibnamefont {Watanabe}}, \bibinfo {author}
  {\bibfnamefont {S.}~\bibnamefont {Katayama}}, \bibinfo {author}
  {\bibfnamefont {H.}~\bibnamefont {Kondo}}, \bibinfo {author} {\bibfnamefont
  {H.}~\bibnamefont {Kitahara}}, \bibinfo {author} {\bibfnamefont
  {S.}~\bibnamefont {Kato}}, \ and\ \bibinfo {author} {\bibfnamefont {M.~W.}\
  \bibnamefont {Takeda}},\ }\href {\doibase 10.1103/PhysRevB.69.245119}
  {\bibfield  {journal} {\bibinfo  {journal} {Phys. Rev. B}\ }\textbf {\bibinfo
  {volume} {69}},\ \bibinfo {pages} {245119} (\bibinfo {year}
  {2004})}\BibitemShut {NoStop}%
\bibitem [{\citenamefont {Ben-Abdallah}\ and\ \citenamefont
  {Ni}(2005)}]{rev1-3}%
  \BibitemOpen
  \bibfield  {author} {\bibinfo {author} {\bibfnamefont {P.}~\bibnamefont
  {Ben-Abdallah}}\ and\ \bibinfo {author} {\bibfnamefont {B.}~\bibnamefont
  {Ni}},\ }\href {\doibase 10.1063/1.1898450} {\bibfield  {journal} {\bibinfo
  {journal} {Journal of Applied Physics}\ }\textbf {\bibinfo {volume} {97}},\
  \bibinfo {pages} {104910} (\bibinfo {year} {2005})},\ \Eprint
  {http://arxiv.org/abs/https://doi.org/10.1063/1.1898450}
  {https://doi.org/10.1063/1.1898450} \BibitemShut {NoStop}%
\bibitem [{\citenamefont {Yu}\ \emph {et~al.}(2019)\citenamefont {Yu},
  \citenamefont {Li},\ and\ \citenamefont {Shen}}]{yu_directional_2019}%
  \BibitemOpen
  \bibfield  {author} {\bibinfo {author} {\bibfnamefont {B.}~\bibnamefont
  {Yu}}, \bibinfo {author} {\bibfnamefont {J.}~\bibnamefont {Li}}, \ and\
  \bibinfo {author} {\bibfnamefont {S.}~\bibnamefont {Shen}},\ }\href {\doibase
  10.1117/1.JPE.9.032712} {\bibfield  {journal} {\bibinfo  {journal} {Journal
  of Photonics for Energy}\ }\textbf {\bibinfo {volume} {9}},\ \bibinfo {pages}
  {032712} (\bibinfo {year} {2019})}\BibitemShut {NoStop}%
\bibitem [{\citenamefont {Hesketh}\ \emph {et~al.}(1988)\citenamefont
  {Hesketh}, \citenamefont {Zemel},\ and\ \citenamefont {Gebhart}}]{rev1-4}%
  \BibitemOpen
  \bibfield  {author} {\bibinfo {author} {\bibfnamefont {P.~J.}\ \bibnamefont
  {Hesketh}}, \bibinfo {author} {\bibfnamefont {J.~N.}\ \bibnamefont {Zemel}},
  \ and\ \bibinfo {author} {\bibfnamefont {B.}~\bibnamefont {Gebhart}},\ }\href
  {\doibase 10.1103/PhysRevB.37.10803} {\bibfield  {journal} {\bibinfo
  {journal} {Phys. Rev. B}\ }\textbf {\bibinfo {volume} {37}},\ \bibinfo
  {pages} {10803} (\bibinfo {year} {1988})}\BibitemShut {NoStop}%
\bibitem [{\citenamefont {Greffet}\ \emph {et~al.}(2002)\citenamefont
  {Greffet}, \citenamefont {Carminati}, \citenamefont {Joulain}, \citenamefont
  {Mulet}, \citenamefont {Mainguy},\ and\ \citenamefont {Chen}}]{rev1-5}%
  \BibitemOpen
  \bibfield  {author} {\bibinfo {author} {\bibfnamefont {J.-J.}\ \bibnamefont
  {Greffet}}, \bibinfo {author} {\bibfnamefont {R.}~\bibnamefont {Carminati}},
  \bibinfo {author} {\bibfnamefont {K.}~\bibnamefont {Joulain}}, \bibinfo
  {author} {\bibfnamefont {J.-P.}\ \bibnamefont {Mulet}}, \bibinfo {author}
  {\bibfnamefont {S.}~\bibnamefont {Mainguy}}, \ and\ \bibinfo {author}
  {\bibfnamefont {Y.}~\bibnamefont {Chen}},\ }\href {\doibase 10.1038/416061a}
  {\bibfield  {journal} {\bibinfo  {journal} {Nature}\ }\textbf {\bibinfo
  {volume} {416}},\ \bibinfo {pages} {61} (\bibinfo {year} {2002})}\BibitemShut
  {NoStop}%
\bibitem [{\citenamefont {Li}\ \emph {et~al.}(2020)\citenamefont {Li},
  \citenamefont {Yu},\ and\ \citenamefont {Shen}}]{li_scale_2020}%
  \BibitemOpen
  \bibfield  {author} {\bibinfo {author} {\bibfnamefont {J.}~\bibnamefont
  {Li}}, \bibinfo {author} {\bibfnamefont {B.}~\bibnamefont {Yu}}, \ and\
  \bibinfo {author} {\bibfnamefont {S.}~\bibnamefont {Shen}},\ }\href {\doibase
  10.1103/PhysRevLett.124.137401} {\bibfield  {journal} {\bibinfo  {journal}
  {Physical Review Letters}\ }\textbf {\bibinfo {volume} {124}},\ \bibinfo
  {pages} {137401} (\bibinfo {year} {2020})}\BibitemShut {NoStop}%
\bibitem [{\citenamefont {Zhu}\ \emph {et~al.}(2013)\citenamefont {Zhu},
  \citenamefont {Sandhu}, \citenamefont {Otey}, \citenamefont {Fan},
  \citenamefont {Sinclair},\ and\ \citenamefont
  {Shan~Luk}}]{zhu_temporal_2013}%
  \BibitemOpen
  \bibfield  {author} {\bibinfo {author} {\bibfnamefont {L.}~\bibnamefont
  {Zhu}}, \bibinfo {author} {\bibfnamefont {S.}~\bibnamefont {Sandhu}},
  \bibinfo {author} {\bibfnamefont {C.}~\bibnamefont {Otey}}, \bibinfo {author}
  {\bibfnamefont {S.}~\bibnamefont {Fan}}, \bibinfo {author} {\bibfnamefont
  {M.~B.}\ \bibnamefont {Sinclair}}, \ and\ \bibinfo {author} {\bibfnamefont
  {T.}~\bibnamefont {Shan~Luk}},\ }\href {\doibase 10.1063/1.4794981}
  {\bibfield  {journal} {\bibinfo  {journal} {Applied Physics Letters}\
  }\textbf {\bibinfo {volume} {102}},\ \bibinfo {pages} {103104} (\bibinfo
  {year} {2013})}\BibitemShut {NoStop}%
\bibitem [{\citenamefont {Koppens}\ \emph {et~al.}(2011)\citenamefont
  {Koppens}, \citenamefont {Chang},\ and\ \citenamefont {García~de
  Abajo}}]{koppens_graphene_2011}%
  \BibitemOpen
  \bibfield  {author} {\bibinfo {author} {\bibfnamefont {F.~H.~L.}\
  \bibnamefont {Koppens}}, \bibinfo {author} {\bibfnamefont {D.~E.}\
  \bibnamefont {Chang}}, \ and\ \bibinfo {author} {\bibfnamefont {F.~J.}\
  \bibnamefont {García~de Abajo}},\ }\href {\doibase 10.1021/nl201771h}
  {\bibfield  {journal} {\bibinfo  {journal} {Nano Letters}\ }\textbf {\bibinfo
  {volume} {11}},\ \bibinfo {pages} {3370} (\bibinfo {year}
  {2011})}\BibitemShut {NoStop}%
\bibitem [{\citenamefont {Yan}\ \emph {et~al.}(2013)\citenamefont {Yan},
  \citenamefont {Low}, \citenamefont {Zhu}, \citenamefont {Wu}, \citenamefont
  {Freitag}, \citenamefont {Li}, \citenamefont {Guinea}, \citenamefont
  {Avouris},\ and\ \citenamefont {Xia}}]{yan_damping_2013}%
  \BibitemOpen
  \bibfield  {author} {\bibinfo {author} {\bibfnamefont {H.}~\bibnamefont
  {Yan}}, \bibinfo {author} {\bibfnamefont {T.}~\bibnamefont {Low}}, \bibinfo
  {author} {\bibfnamefont {W.}~\bibnamefont {Zhu}}, \bibinfo {author}
  {\bibfnamefont {Y.}~\bibnamefont {Wu}}, \bibinfo {author} {\bibfnamefont
  {M.}~\bibnamefont {Freitag}}, \bibinfo {author} {\bibfnamefont
  {X.}~\bibnamefont {Li}}, \bibinfo {author} {\bibfnamefont {F.}~\bibnamefont
  {Guinea}}, \bibinfo {author} {\bibfnamefont {P.}~\bibnamefont {Avouris}}, \
  and\ \bibinfo {author} {\bibfnamefont {F.}~\bibnamefont {Xia}},\ }\href
  {\doibase 10.1038/nphoton.2013.57} {\bibfield  {journal} {\bibinfo  {journal}
  {Nature Photonics}\ }\textbf {\bibinfo {volume} {7}},\ \bibinfo {pages} {394}
  (\bibinfo {year} {2013})}\BibitemShut {NoStop}%
\bibitem [{\citenamefont {Brar}\ \emph {et~al.}(2015)\citenamefont {Brar},
  \citenamefont {Sherrott}, \citenamefont {Jang}, \citenamefont {Kim},
  \citenamefont {Kim}, \citenamefont {Choi}, \citenamefont {Sweatlock},\ and\
  \citenamefont {Atwater}}]{brar_electronic_2015}%
  \BibitemOpen
  \bibfield  {author} {\bibinfo {author} {\bibfnamefont {V.~W.}\ \bibnamefont
  {Brar}}, \bibinfo {author} {\bibfnamefont {M.~C.}\ \bibnamefont {Sherrott}},
  \bibinfo {author} {\bibfnamefont {M.~S.}\ \bibnamefont {Jang}}, \bibinfo
  {author} {\bibfnamefont {S.}~\bibnamefont {Kim}}, \bibinfo {author}
  {\bibfnamefont {L.}~\bibnamefont {Kim}}, \bibinfo {author} {\bibfnamefont
  {M.}~\bibnamefont {Choi}}, \bibinfo {author} {\bibfnamefont {L.~A.}\
  \bibnamefont {Sweatlock}}, \ and\ \bibinfo {author} {\bibfnamefont {H.~A.}\
  \bibnamefont {Atwater}},\ }\href {\doibase 10.1038/ncomms8032} {\bibfield
  {journal} {\bibinfo  {journal} {Nature Communications}\ }\textbf {\bibinfo
  {volume} {6}},\ \bibinfo {pages} {7032} (\bibinfo {year} {2015})}\BibitemShut
  {NoStop}%
\bibitem [{\citenamefont {Guo}\ \emph {et~al.}(2017)\citenamefont {Guo},
  \citenamefont {Li}, \citenamefont {Deng}, \citenamefont {Yuan}, \citenamefont
  {Guinea},\ and\ \citenamefont {Xia}}]{guo_infrared_2017}%
  \BibitemOpen
  \bibfield  {author} {\bibinfo {author} {\bibfnamefont {Q.}~\bibnamefont
  {Guo}}, \bibinfo {author} {\bibfnamefont {C.}~\bibnamefont {Li}}, \bibinfo
  {author} {\bibfnamefont {B.}~\bibnamefont {Deng}}, \bibinfo {author}
  {\bibfnamefont {S.}~\bibnamefont {Yuan}}, \bibinfo {author} {\bibfnamefont
  {F.}~\bibnamefont {Guinea}}, \ and\ \bibinfo {author} {\bibfnamefont
  {F.}~\bibnamefont {Xia}},\ }\href {\doibase 10.1021/acsphotonics.7b00547}
  {\bibfield  {journal} {\bibinfo  {journal} {ACS Photonics}\ }\textbf
  {\bibinfo {volume} {4}},\ \bibinfo {pages} {2989} (\bibinfo {year}
  {2017})}\BibitemShut {NoStop}%
\bibitem [{\citenamefont {Jablan}\ \emph {et~al.}(2009)\citenamefont {Jablan},
  \citenamefont {Buljan},\ and\ \citenamefont
  {Soljacic}}]{jablan_plasmonics_2009}%
  \BibitemOpen
  \bibfield  {author} {\bibinfo {author} {\bibfnamefont {M.}~\bibnamefont
  {Jablan}}, \bibinfo {author} {\bibfnamefont {H.}~\bibnamefont {Buljan}}, \
  and\ \bibinfo {author} {\bibfnamefont {M.}~\bibnamefont {Soljacic}},\
  }\href {\doibase 10.1103/PhysRevB.80.245435} {\bibfield  {journal} {\bibinfo
  {journal} {Physical Review B}\ }\textbf {\bibinfo {volume} {80}},\ \bibinfo
  {pages} {245435} (\bibinfo {year} {2009})}\BibitemShut {NoStop}%
\bibitem [{\citenamefont {Chen}\ \emph {et~al.}(2011)\citenamefont {Chen},
  \citenamefont {Park}, \citenamefont {Boudouris}, \citenamefont {Horng},
  \citenamefont {Geng}, \citenamefont {Girit}, \citenamefont {Zettl},
  \citenamefont {Crommie}, \citenamefont {Segalman}, \citenamefont {Louie},\
  and\ \citenamefont {Wang}}]{chen_controlling_2011}%
  \BibitemOpen
  \bibfield  {author} {\bibinfo {author} {\bibfnamefont {C.-F.}\ \bibnamefont
  {Chen}}, \bibinfo {author} {\bibfnamefont {C.-H.}\ \bibnamefont {Park}},
  \bibinfo {author} {\bibfnamefont {B.~W.}\ \bibnamefont {Boudouris}}, \bibinfo
  {author} {\bibfnamefont {J.}~\bibnamefont {Horng}}, \bibinfo {author}
  {\bibfnamefont {B.}~\bibnamefont {Geng}}, \bibinfo {author} {\bibfnamefont
  {C.}~\bibnamefont {Girit}}, \bibinfo {author} {\bibfnamefont
  {A.}~\bibnamefont {Zettl}}, \bibinfo {author} {\bibfnamefont {M.~F.}\
  \bibnamefont {Crommie}}, \bibinfo {author} {\bibfnamefont {R.~A.}\
  \bibnamefont {Segalman}}, \bibinfo {author} {\bibfnamefont {S.~G.}\
  \bibnamefont {Louie}}, \ and\ \bibinfo {author} {\bibfnamefont
  {F.}~\bibnamefont {Wang}},\ }\href {\doibase 10.1038/nature09866} {\bibfield
  {journal} {\bibinfo  {journal} {Nature}\ }\textbf {\bibinfo {volume} {471}},\
  \bibinfo {pages} {617} (\bibinfo {year} {2011})}\BibitemShut {NoStop}%
\bibitem [{\citenamefont {Efetov}\ and\ \citenamefont
  {Kim}(2010)}]{efetov_controlling_2010}%
  \BibitemOpen
  \bibfield  {author} {\bibinfo {author} {\bibfnamefont {D.~K.}\ \bibnamefont
  {Efetov}}\ and\ \bibinfo {author} {\bibfnamefont {P.}~\bibnamefont {Kim}},\
  }\href {\doibase 10.1103/PhysRevLett.105.256805} {\bibfield  {journal}
  {\bibinfo  {journal} {Physical Review Letters}\ }\textbf {\bibinfo {volume}
  {105}},\ \bibinfo {pages} {256805} (\bibinfo {year} {2010})}\BibitemShut
  {NoStop}%
\bibitem [{\citenamefont {Zhao}\ \emph {et~al.}(2015)\citenamefont {Zhao},
  \citenamefont {Zhao},\ and\ \citenamefont {Zhang}}]{rev2-1}%
  \BibitemOpen
  \bibfield  {author} {\bibinfo {author} {\bibfnamefont {B.}~\bibnamefont
  {Zhao}}, \bibinfo {author} {\bibfnamefont {J.~M.}\ \bibnamefont {Zhao}}, \
  and\ \bibinfo {author} {\bibfnamefont {Z.~M.}\ \bibnamefont {Zhang}},\ }\href
  {\doibase 10.1364/JOSAB.32.001176} {\bibfield  {journal} {\bibinfo  {journal}
  {J. Opt. Soc. Am. B}\ }\textbf {\bibinfo {volume} {32}},\ \bibinfo {pages}
  {1176} (\bibinfo {year} {2015})}\BibitemShut {NoStop}%
\bibitem [{\citenamefont {Pan}\ \emph {et~al.}(2017)\citenamefont {Pan},
  \citenamefont {Hong}, \citenamefont {Zhang}, \citenamefont {Shuai},\ and\
  \citenamefont {Tan}}]{rev2-2}%
  \BibitemOpen
  \bibfield  {author} {\bibinfo {author} {\bibfnamefont {Q.}~\bibnamefont
  {Pan}}, \bibinfo {author} {\bibfnamefont {J.}~\bibnamefont {Hong}}, \bibinfo
  {author} {\bibfnamefont {G.}~\bibnamefont {Zhang}}, \bibinfo {author}
  {\bibfnamefont {Y.}~\bibnamefont {Shuai}}, \ and\ \bibinfo {author}
  {\bibfnamefont {H.}~\bibnamefont {Tan}},\ }\href {\doibase
  10.1364/OE.25.016400} {\bibfield  {journal} {\bibinfo  {journal} {Opt.
  Express}\ }\textbf {\bibinfo {volume} {25}},\ \bibinfo {pages} {16400}
  (\bibinfo {year} {2017})}\BibitemShut {NoStop}%
\bibitem [{\citenamefont {Liu}\ \emph {et~al.}(2015)\citenamefont {Liu},
  \citenamefont {Zhao},\ and\ \citenamefont {Zhang}}]{rev2-3}%
  \BibitemOpen
  \bibfield  {author} {\bibinfo {author} {\bibfnamefont {X.~L.}\ \bibnamefont
  {Liu}}, \bibinfo {author} {\bibfnamefont {B.}~\bibnamefont {Zhao}}, \ and\
  \bibinfo {author} {\bibfnamefont {Z.~M.}\ \bibnamefont {Zhang}},\ }\href
  {\doibase 10.1088/2040-8978/17/3/035004} {\bibfield  {journal} {\bibinfo
  {journal} {Journal of Optics}\ }\textbf {\bibinfo {volume} {17}},\ \bibinfo
  {pages} {035004} (\bibinfo {year} {2015})}\BibitemShut {NoStop}%
\bibitem [{\citenamefont {Wang}\ \emph {et~al.}(2015)\citenamefont {Wang},
  \citenamefont {Yang},\ and\ \citenamefont {Wang}}]{rev2-4}%
  \BibitemOpen
  \bibfield  {author} {\bibinfo {author} {\bibfnamefont {H.}~\bibnamefont
  {Wang}}, \bibinfo {author} {\bibfnamefont {Y.}~\bibnamefont {Yang}}, \ and\
  \bibinfo {author} {\bibfnamefont {L.}~\bibnamefont {Wang}},\ }\href {\doibase
  10.1088/2040-8978/17/4/045104} {\bibfield  {journal} {\bibinfo  {journal}
  {Journal of Optics}\ }\textbf {\bibinfo {volume} {17}},\ \bibinfo {pages}
  {045104} (\bibinfo {year} {2015})}\BibitemShut {NoStop}%
\bibitem [{\citenamefont {Lin}\ \emph {et~al.}(2020)\citenamefont {Lin},
  \citenamefont {Lin}, \citenamefont {Yang},\ and\ \citenamefont
  {Jia}}]{rev2-5}%
  \BibitemOpen
  \bibfield  {author} {\bibinfo {author} {\bibfnamefont {K.-T.}\ \bibnamefont
  {Lin}}, \bibinfo {author} {\bibfnamefont {H.}~\bibnamefont {Lin}}, \bibinfo
  {author} {\bibfnamefont {T.}~\bibnamefont {Yang}}, \ and\ \bibinfo {author}
  {\bibfnamefont {B.}~\bibnamefont {Jia}},\ }\href {\doibase
  10.1038/s41467-020-15116-z} {\bibfield  {journal} {\bibinfo  {journal}
  {Nature Communications}\ }\textbf {\bibinfo {volume} {11}},\ \bibinfo {pages}
  {1389} (\bibinfo {year} {2020})}\BibitemShut {NoStop}%
\bibitem [{\citenamefont {Cai}\ \emph {et~al.}(2015{\natexlab{a}})\citenamefont
  {Cai}, \citenamefont {Zhu}, \citenamefont {Liu}, \citenamefont {Lin},
  \citenamefont {Zhou}, \citenamefont {Ye},\ and\ \citenamefont
  {Cai}}]{rev2-6}%
  \BibitemOpen
  \bibfield  {author} {\bibinfo {author} {\bibfnamefont {Y.}~\bibnamefont
  {Cai}}, \bibinfo {author} {\bibfnamefont {J.}~\bibnamefont {Zhu}}, \bibinfo
  {author} {\bibfnamefont {Q.~H.}\ \bibnamefont {Liu}}, \bibinfo {author}
  {\bibfnamefont {T.}~\bibnamefont {Lin}}, \bibinfo {author} {\bibfnamefont
  {J.}~\bibnamefont {Zhou}}, \bibinfo {author} {\bibfnamefont {L.}~\bibnamefont
  {Ye}}, \ and\ \bibinfo {author} {\bibfnamefont {Z.}~\bibnamefont {Cai}},\
  }\href {\doibase 10.1364/OE.23.032318} {\bibfield  {journal} {\bibinfo
  {journal} {Opt. Express}\ }\textbf {\bibinfo {volume} {23}},\ \bibinfo
  {pages} {32318} (\bibinfo {year} {2015}{\natexlab{a}})}\BibitemShut {NoStop}%
\bibitem [{\citenamefont {Cai}\ \emph {et~al.}(2015{\natexlab{b}})\citenamefont
  {Cai}, \citenamefont {Zhu},\ and\ \citenamefont {Liu}}]{rev2-7}%
  \BibitemOpen
  \bibfield  {author} {\bibinfo {author} {\bibfnamefont {Y.}~\bibnamefont
  {Cai}}, \bibinfo {author} {\bibfnamefont {J.}~\bibnamefont {Zhu}}, \ and\
  \bibinfo {author} {\bibfnamefont {Q.~H.}\ \bibnamefont {Liu}},\ }\href
  {\doibase 10.1063/1.4906996} {\bibfield  {journal} {\bibinfo  {journal}
  {Applied Physics Letters}\ }\textbf {\bibinfo {volume} {106}},\ \bibinfo
  {pages} {043105} (\bibinfo {year} {2015}{\natexlab{b}})},\ \Eprint
  {http://arxiv.org/abs/https://doi.org/10.1063/1.4906996}
  {https://doi.org/10.1063/1.4906996} \BibitemShut {NoStop}%
\bibitem [{\citenamefont {Xiao}\ \emph {et~al.}(2019)\citenamefont {Xiao},
  \citenamefont {Liu}, \citenamefont {Lei}, \citenamefont {Sun}, \citenamefont
  {Ouyang},\ and\ \citenamefont {Tao}}]{rev2-8}%
  \BibitemOpen
  \bibfield  {author} {\bibinfo {author} {\bibfnamefont {D.}~\bibnamefont
  {Xiao}}, \bibinfo {author} {\bibfnamefont {Q.}~\bibnamefont {Liu}}, \bibinfo
  {author} {\bibfnamefont {L.}~\bibnamefont {Lei}}, \bibinfo {author}
  {\bibfnamefont {Y.}~\bibnamefont {Sun}}, \bibinfo {author} {\bibfnamefont
  {Z.}~\bibnamefont {Ouyang}}, \ and\ \bibinfo {author} {\bibfnamefont
  {K.}~\bibnamefont {Tao}},\ }\href {\doibase 10.1186/s11671-019-2852-y}
  {\bibfield  {journal} {\bibinfo  {journal} {Nanoscale Research Letters}\
  }\textbf {\bibinfo {volume} {14}},\ \bibinfo {pages} {32} (\bibinfo {year}
  {2019})},\ \Eprint
  {http://arxiv.org/abs/https://doi.org/10.1186/s11671-019-2852-y}
  {https://doi.org/10.1186/s11671-019-2852-y} \BibitemShut {NoStop}%
\bibitem [{\citenamefont {Li}\ \emph {et~al.}(2017)\citenamefont {Li},
  \citenamefont {Liu},\ and\ \citenamefont {Shen}}]{li_tunable_2017}%
  \BibitemOpen
  \bibfield  {author} {\bibinfo {author} {\bibfnamefont {J.}~\bibnamefont
  {Li}}, \bibinfo {author} {\bibfnamefont {B.}~\bibnamefont {Liu}}, \ and\
  \bibinfo {author} {\bibfnamefont {S.}~\bibnamefont {Shen}},\ }\href {\doibase
  10.1103/PhysRevB.96.075413} {\bibfield  {journal} {\bibinfo  {journal}
  {Physical Review B}\ }\textbf {\bibinfo {volume} {96}},\ \bibinfo {pages}
  {075413} (\bibinfo {year} {2017})}\BibitemShut {NoStop}%
\bibitem [{\citenamefont {Li}\ \emph {et~al.}(2021)\citenamefont {Li},
  \citenamefont {Li}, \citenamefont {Liu}, \citenamefont {Salihoglu},\ and\
  \citenamefont {Shen}}]{li_wiener_2021}%
  \BibitemOpen
  \bibfield  {author} {\bibinfo {author} {\bibfnamefont {Z.}~\bibnamefont
  {Li}}, \bibinfo {author} {\bibfnamefont {J.}~\bibnamefont {Li}}, \bibinfo
  {author} {\bibfnamefont {X.}~\bibnamefont {Liu}}, \bibinfo {author}
  {\bibfnamefont {H.}~\bibnamefont {Salihoglu}}, \ and\ \bibinfo {author}
  {\bibfnamefont {S.}~\bibnamefont {Shen}},\ }\href {\doibase
  10.1103/PhysRevB.104.195426} {\bibfield  {journal} {\bibinfo  {journal}
  {Physical Review B}\ }\textbf {\bibinfo {volume} {104}},\ \bibinfo {pages}
  {195426} (\bibinfo {year} {2021})}\BibitemShut {NoStop}%
\bibitem [{\citenamefont {Hanson}(2008)}]{graphene_mat}%
  \BibitemOpen
  \bibfield  {author} {\bibinfo {author} {\bibfnamefont {G.~W.}\ \bibnamefont
  {Hanson}},\ }\href {\doibase 10.1063/1.2891452} {\bibfield  {journal}
  {\bibinfo  {journal} {Journal of Applied Physics}\ }\textbf {\bibinfo
  {volume} {103}},\ \bibinfo {pages} {064302} (\bibinfo {year} {2008})},\
  \Eprint {http://arxiv.org/abs/https://doi.org/10.1063/1.2891452}
  {https://doi.org/10.1063/1.2891452} \BibitemShut {NoStop}%
\bibitem [{\citenamefont {Wood}\ \emph {et~al.}(1990)\citenamefont {Wood},
  \citenamefont {Nassau}, \citenamefont {Kometani},\ and\ \citenamefont
  {Nash}}]{HfO2_mat}%
  \BibitemOpen
  \bibfield  {author} {\bibinfo {author} {\bibfnamefont {D.~L.}\ \bibnamefont
  {Wood}}, \bibinfo {author} {\bibfnamefont {K.}~\bibnamefont {Nassau}},
  \bibinfo {author} {\bibfnamefont {T.~Y.}\ \bibnamefont {Kometani}}, \ and\
  \bibinfo {author} {\bibfnamefont {D.~L.}\ \bibnamefont {Nash}},\ }\href
  {\doibase 10.1364/AO.29.000604} {\bibfield  {journal} {\bibinfo  {journal}
  {Appl. Opt.}\ }\textbf {\bibinfo {volume} {29}},\ \bibinfo {pages} {604}
  (\bibinfo {year} {1990})}\BibitemShut {NoStop}%
\bibitem [{\citenamefont {Ordal}\ \emph {et~al.}(1985)\citenamefont {Ordal},
  \citenamefont {Bell}, \citenamefont {Alexander}, \citenamefont {Long},\ and\
  \citenamefont {Querry}}]{Au_mat}%
  \BibitemOpen
  \bibfield  {author} {\bibinfo {author} {\bibfnamefont {M.~A.}\ \bibnamefont
  {Ordal}}, \bibinfo {author} {\bibfnamefont {R.~J.}\ \bibnamefont {Bell}},
  \bibinfo {author} {\bibfnamefont {R.~W.}\ \bibnamefont {Alexander}}, \bibinfo
  {author} {\bibfnamefont {L.~L.}\ \bibnamefont {Long}}, \ and\ \bibinfo
  {author} {\bibfnamefont {M.~R.}\ \bibnamefont {Querry}},\ }\href {\doibase
  10.1364/AO.24.004493} {\bibfield  {journal} {\bibinfo  {journal} {Appl.
  Opt.}\ }\textbf {\bibinfo {volume} {24}},\ \bibinfo {pages} {4493} (\bibinfo
  {year} {1985})}\BibitemShut {NoStop}%
\bibitem [{\citenamefont {edited~by Edward D.~Palik}(1985)}]{Palik}%
  \BibitemOpen
  \bibfield  {author} {\bibinfo {author} {\bibnamefont {edited~by Edward
  D.~Palik}},\ }\href {https://search.library.wisc.edu/catalog/999554063402121}
  {\emph {\bibinfo {title} {Handbook of optical constants of solids}}}\
  (\bibinfo  {publisher} {Orlando : Academic Press, 1985.},\ \bibinfo {year}
  {1985})\ \bibinfo {note} {includes bibliographies.}\BibitemShut {Stop}%
\bibitem [{\citenamefont {Liu}\ \emph {et~al.}(2014)\citenamefont {Liu},
  \citenamefont {Shi}, \citenamefont {Liew},\ and\ \citenamefont
  {Shen}}]{Si_mat_1}%
  \BibitemOpen
  \bibfield  {author} {\bibinfo {author} {\bibfnamefont {B.}~\bibnamefont
  {Liu}}, \bibinfo {author} {\bibfnamefont {J.}~\bibnamefont {Shi}}, \bibinfo
  {author} {\bibfnamefont {K.}~\bibnamefont {Liew}}, \ and\ \bibinfo {author}
  {\bibfnamefont {S.}~\bibnamefont {Shen}},\ }\href {\doibase
  https://doi.org/10.1016/j.optcom.2013.10.074} {\bibfield  {journal} {\bibinfo
   {journal} {Optics Communications}\ }\textbf {\bibinfo {volume} {314}},\
  \bibinfo {pages} {57} (\bibinfo {year} {2014})},\ \bibinfo {note} {energy
  efficient nanophotonics: Engineered light–matter interaction in
  sub-wavelength structures}\BibitemShut {NoStop}%
\bibitem [{\citenamefont {Fu}\ and\ \citenamefont {Zhang}(2006)}]{Si_mat_2}%
  \BibitemOpen
  \bibfield  {author} {\bibinfo {author} {\bibfnamefont {C.}~\bibnamefont
  {Fu}}\ and\ \bibinfo {author} {\bibfnamefont {Z.}~\bibnamefont {Zhang}},\
  }\href {\doibase https://doi.org/10.1016/j.ijheatmasstransfer.2005.09.037}
  {\bibfield  {journal} {\bibinfo  {journal} {International Journal of Heat and
  Mass Transfer}\ }\textbf {\bibinfo {volume} {49}},\ \bibinfo {pages} {1703}
  (\bibinfo {year} {2006})}\BibitemShut {NoStop}%
\bibitem [{\citenamefont {Basu}\ \emph {et~al.}(2009)\citenamefont {Basu},
  \citenamefont {Zhang},\ and\ \citenamefont {Fu}}]{Si_mat_3}%
  \BibitemOpen
  \bibfield  {author} {\bibinfo {author} {\bibfnamefont {S.}~\bibnamefont
  {Basu}}, \bibinfo {author} {\bibfnamefont {Z.~M.}\ \bibnamefont {Zhang}}, \
  and\ \bibinfo {author} {\bibfnamefont {C.~J.}\ \bibnamefont {Fu}},\ }\href
  {\doibase https://doi.org/10.1002/er.1607} {\bibfield  {journal} {\bibinfo
  {journal} {International Journal of Energy Research}\ }\textbf {\bibinfo
  {volume} {33}},\ \bibinfo {pages} {1203} (\bibinfo {year} {2009})},\ \Eprint
  {http://arxiv.org/abs/https://onlinelibrary.wiley.com/doi/pdf/10.1002/er.1607}
  {https://onlinelibrary.wiley.com/doi/pdf/10.1002/er.1607} \BibitemShut
  {NoStop}%
\bibitem [{\citenamefont {Fan}\ \emph {et~al.}(2016)\citenamefont {Fan},
  \citenamefont {Suen}, \citenamefont {Wu},\ and\ \citenamefont
  {Padilla}}]{xiu-1}%
  \BibitemOpen
  \bibfield  {author} {\bibinfo {author} {\bibfnamefont {K.}~\bibnamefont
  {Fan}}, \bibinfo {author} {\bibfnamefont {J.}~\bibnamefont {Suen}}, \bibinfo
  {author} {\bibfnamefont {X.}~\bibnamefont {Wu}}, \ and\ \bibinfo {author}
  {\bibfnamefont {W.~J.}\ \bibnamefont {Padilla}},\ }\href {\doibase
  10.1364/OE.24.025189} {\bibfield  {journal} {\bibinfo  {journal} {Opt.
  Express}\ }\textbf {\bibinfo {volume} {24}},\ \bibinfo {pages} {25189}
  (\bibinfo {year} {2016})}\BibitemShut {NoStop}%
\bibitem [{\citenamefont {Zeng}\ \emph {et~al.}(2018)\citenamefont {Zeng},
  \citenamefont {Huang}, \citenamefont {Singh}, \citenamefont {Yao},
  \citenamefont {Azad}, \citenamefont {Mohite}, \citenamefont {Taylor},
  \citenamefont {Smith},\ and\ \citenamefont {Chen}}]{xiu-2}%
  \BibitemOpen
  \bibfield  {author} {\bibinfo {author} {\bibfnamefont {B.}~\bibnamefont
  {Zeng}}, \bibinfo {author} {\bibfnamefont {Z.}~\bibnamefont {Huang}},
  \bibinfo {author} {\bibfnamefont {A.}~\bibnamefont {Singh}}, \bibinfo
  {author} {\bibfnamefont {Y.}~\bibnamefont {Yao}}, \bibinfo {author}
  {\bibfnamefont {A.~K.}\ \bibnamefont {Azad}}, \bibinfo {author}
  {\bibfnamefont {A.~D.}\ \bibnamefont {Mohite}}, \bibinfo {author}
  {\bibfnamefont {A.~J.}\ \bibnamefont {Taylor}}, \bibinfo {author}
  {\bibfnamefont {D.~R.}\ \bibnamefont {Smith}}, \ and\ \bibinfo {author}
  {\bibfnamefont {H.-T.}\ \bibnamefont {Chen}},\ }\href {\doibase
  10.1038/s41377-018-0055-4} {\bibfield  {journal} {\bibinfo  {journal} {Light:
  Science \& Applications}\ }\textbf {\bibinfo {volume} {7}},\ \bibinfo {pages}
  {51} (\bibinfo {year} {2018})}\BibitemShut {NoStop}%
\bibitem [{\citenamefont {Yao}\ \emph {et~al.}(2014)\citenamefont {Yao},
  \citenamefont {Shankar}, \citenamefont {Kats}, \citenamefont {Song},
  \citenamefont {Kong}, \citenamefont {Loncar},\ and\ \citenamefont
  {Capasso}}]{xiu-3}%
  \BibitemOpen
  \bibfield  {author} {\bibinfo {author} {\bibfnamefont {Y.}~\bibnamefont
  {Yao}}, \bibinfo {author} {\bibfnamefont {R.}~\bibnamefont {Shankar}},
  \bibinfo {author} {\bibfnamefont {M.~A.}\ \bibnamefont {Kats}}, \bibinfo
  {author} {\bibfnamefont {Y.}~\bibnamefont {Song}}, \bibinfo {author}
  {\bibfnamefont {J.}~\bibnamefont {Kong}}, \bibinfo {author} {\bibfnamefont
  {M.}~\bibnamefont {Loncar}}, \ and\ \bibinfo {author} {\bibfnamefont
  {F.}~\bibnamefont {Capasso}},\ }\href {\doibase 10.1021/nl503104n} {\bibfield
   {journal} {\bibinfo  {journal} {Nano Letters}\ }\textbf {\bibinfo {volume}
  {14}},\ \bibinfo {pages} {6526} (\bibinfo {year} {2014})},\ \bibinfo {note}
  {pMID: 25310847},\ \Eprint
  {http://arxiv.org/abs/https://doi.org/10.1021/nl503104n}
  {https://doi.org/10.1021/nl503104n} \BibitemShut {NoStop}%
\bibitem [{\citenamefont {Wojszvzyk}\ \emph {et~al.}(2021)\citenamefont
  {Wojszvzyk}, \citenamefont {Nguyen}, \citenamefont {Coutrot}, \citenamefont
  {Zhang}, \citenamefont {Vest},\ and\ \citenamefont {Greffet}}]{xiu-4}%
  \BibitemOpen
  \bibfield  {author} {\bibinfo {author} {\bibfnamefont {L.}~\bibnamefont
  {Wojszvzyk}}, \bibinfo {author} {\bibfnamefont {A.}~\bibnamefont {Nguyen}},
  \bibinfo {author} {\bibfnamefont {A.-L.}\ \bibnamefont {Coutrot}}, \bibinfo
  {author} {\bibfnamefont {C.}~\bibnamefont {Zhang}}, \bibinfo {author}
  {\bibfnamefont {B.}~\bibnamefont {Vest}}, \ and\ \bibinfo {author}
  {\bibfnamefont {J.-J.}\ \bibnamefont {Greffet}},\ }\href {\doibase
  10.1038/s41467-021-21752-w} {\bibfield  {journal} {\bibinfo  {journal}
  {Nature Communications}\ }\textbf {\bibinfo {volume} {12}},\ \bibinfo {pages}
  {1492} (\bibinfo {year} {2021})}\BibitemShut {NoStop}%
\bibitem [{Note1()}]{Note1}%
  \BibitemOpen
  \bibinfo {note} {The data that support the findings of this study are
  available from the corresponding author upon request}\BibitemShut {NoStop}%
\end{thebibliography}

%

\end{document}